\documentclass[sigconf, screen]{acmart}
\AtBeginDocument{%
  }
\setcopyright{none} 
\copyrightyear{2025}
\acmYear{2025}
\usepackage{algorithm}
\usepackage[noend]{algpseudocode}
\usepackage{graphicx}
\usepackage{multirow}
\usepackage{xspace}
\usepackage{enumitem}
\usepackage[many]{tcolorbox}
\usepackage{wrapfig,lipsum,booktabs}
\usepackage{fancyhdr}

\usepackage{textcomp}
 
\newcommand{\NAME}{$\mathtt{ObfusBFA}$\xspace}    
\renewcommand\footnotetextcopyrightpermission[1]{} 
\begin{document}

\title{\NAME: A Holistic Approach to Safeguarding DNNs from Different Types of Bit-Flip Attacks}
\author{Xiaobei Yan}
\affiliation{\institution{Nanyang Technological University}\country{Singapore}}
\email{xiaobei002@e.ntu.edu.sg}
\author{Han Qiu}
\affiliation{\institution{Tsinghua University}\country{China}}
\email{qiuhan@tsinghua.edu.cn}
\author{Tianwei Zhang}
\affiliation{\institution{Nanyang Technological University}\country{Singapore}}
\email{tianwei.zhang@ntu.edu.sg}

\begin{abstract}
Bit-flip attacks (BFAs) represent a serious threat to Deep Neural Networks (DNNs), where flipping a small number of bits in the model parameters or binary code can significantly degrade the model accuracy or mislead the model prediction in a desired way. Existing defenses exclusively focus on protecting models for specific attacks and platforms, while lacking effectiveness for other scenarios. We propose \NAME, an efficient and holistic methodology to mitigate BFAs targeting both the high-level model weights and low-level codebase (executables or shared libraries). The key idea of \NAME is to introduce random dummy operations during the model inference, which effectively transforms the delicate attacks into random bit flips, making it much harder for attackers to pinpoint and exploit vulnerable bits. We design novel algorithms to identify critical bits and insert obfuscation operations. We evaluate \NAME against different types of attacks, including the adaptive scenarios where the attacker increases the flip bit budget to attempt to circumvent our defense. The results show that \NAME can consistently preserve the model accuracy across various datasets and DNN architectures while significantly reducing the attack success rates. Additionally, it introduces minimal latency and storage overhead, making it a practical solution for real-world applications.
\end{abstract}

\maketitle
\section{Introduction}
\label{sec:Introduction}
The rapid growth of deep learning technology in recent years has driven the widespread deployment of Deep Neural Network (DNN) models across a variety of computing platforms, from autonomous driving~\cite{sato2021dirty} to embedded devices~\cite{zhang2021optimizing}. However, the proliferation of DNNs has exposed new security vulnerabilities. Past studies have shown that modern hardware platforms are susceptible to Bit Flip Attacks (BFAs) \cite{liu2017fault}, which corrupt the critical data or code of the applications, significantly compromising their integrity. Such attacks could be executed through different fault injection techniques, such as row hammering~\cite{mutlu2019rowhammer,yao2020deephammer}, undervolting~\cite{rakin2021deep}, and laser beaming~\cite{breier2018practical}. In the domain of deep learning, researchers apply BFAs to flip a small number of critical bits in the DNN applications \cite{yao2020deephammer,rakin2019bit}, which can severely degrade the model's overall accuracy, or misprediction for certain input. Such attacks can lead to catastrophic consequences, especially in safety-critical scenarios.

\begin{figure}[t]
    \centering
    \includegraphics[scale=0.5]{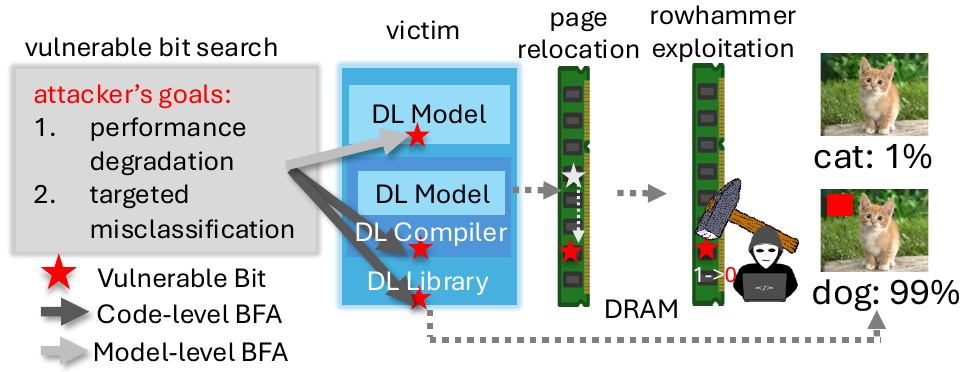}
    \caption{BFAs against DNN applications with Rowhammer.}
    \label{fig:bfaalevels}
\end{figure}

Existing BFAs against DNN applications can be classified into two categories, as demonstrated in Figure \ref{fig:bfaalevels}. 
(1) \textit{Model-level attacks} \cite{chen2021proflip,rakin2019bit,rakin2020tbt,yao2020deephammer,rakin2021t,bai2021targeted}: the attacker compromises some critical model parameters. Even flipping a tiny number of parameter bits can drastically alter the model behaviors, resulting in a sharp drop in accuracy or misprediction in the attacker's desired manner. 
(2) \textit{Code-level attacks} \cite{li2024yes, chen2023unveiling}: the attacker targets the functional code of the deep learning framework, including the executables generated by the compilers, or the underlying computation libraries. By flipping critical bits in these foundational components, the attacker can hijack the control flow of the inference execution, leading to large performance degradation or even system failure. 

\subsection{Existing Studies and Limitations}
\label{sec:exist-study}
In spite of existing efforts to defend DNN models against BFAs, it is still challenging to design a holistic solution to provide comprehensive protection over various threat scenarios. As BFAs are commonly realized via exploiting certain hardware vulnerabilities, a straightforward strategy is to fundamentally mitigate those threats. For instance, a number of solutions have been introduced to defeat Rowhammer attacks, including the adoption of timers~\cite{irazoqui2016mascat}, performance counters~\cite{gruss2016flush+,chiappetta2016real}, error-correcting codes~\cite{cojocar2019exploiting}, memory isolation~\cite{konoth2018zebram, loughlin2023siloz}, or machine learning-based detection~\cite{joardar2022machine}. While effective, these solutions often require specialized hardware upgrades, which can introduce compatibility issues in existing systems and limit their broader adoption. Besides, they lack generalization across different computing platforms and scenarios.

Due to the above limitations, researchers have been actively seeking for new solutions dedicated to the protection of DNN models against BFAs. Existing approaches can be roughly classified into two categories. (1) \textit{Redesigning the DNN model for better robustness}. This is normally achieved via model architecture modifications or retraining. Typical examples include the dynamic multi-exit architecture~\cite{wang2023aegis}, weight reconstruction~\cite{li2020defending}, quantization~\cite{he2020defending,rakin2021ra}, obfuscations of channels~\cite{luo2024deepshuffle,gongye2023hammerdodger} or bits~\cite{liu2022generating}. (2) \textit{Runtime integrity checking}. This strategy continuously monitors the DNN inference, detects and corrects potential errors caused by flipped bits. Some works introduce checksum~\cite{li2021radar,chen2bitshield}, hash functions~\cite{javaheripi2022acchashtag,javaheripi2021hashtag}, or checker DNNs~\cite{li2020deepdyve} for integrity verification. Other studies focus on the detection of critical bits as BFA targets~\cite{hosseini2021safeguarding,liu2023neuropots,liu2020concurrent}.

Unfortunately, these approaches still exhibit several drawbacks in practice. 
(1) \textbf{High access requirement to the DNN models}. Some solutions require significant changes over the model parameters or network designs \cite{wang2023aegis,li2020defending,he2020defending,rakin2021ra,luo2024deepshuffle,gongye2023hammerdodger,liu2022generating}, which are often difficult for pre-deployed models or legacy systems. Some solutions require the access to the original training data~\cite{chen2bitshield}, which may not be always feasible. 
(2) \textbf{Non-negligible runtime overhead}. Runtime integrity checking incurs additional latency for the inference execution, making it impractical for applications with strong real-time requirements~\cite{li2020deepdyve,liu2023neuropots}. For example, on the CIFAR-10 dataset, DeepDyve~\cite{li2020deepdyve} incurs a 9.1\% computational overhead, while NeuroPots~\cite{liu2023neuropots} adds a 9.7\% time overhead.
(3) \textbf{Degradation of model functionality}. Solutions that modify model parameters often compromise model performance while providing protection~\cite{liu2023neuropots,chen2019deepattest,wang2023aegis,he2020defending,rakin2021ra}. As reported in prior work~\cite{chen2bitshield,wang2023aegis}, BIN~\cite{rakin2021ra} reduces model accuracy by approximately 4.1\%, RA-BNN~\cite{rakin2021ra} by 2.9\%, NeuroPots~\cite{liu2023neuropots} by 1.4\%, and Aegis~\cite{wang2023aegis} by 1.2\%. 
(4) \textbf{Lack of generalization}. More importantly, these solutions only target one specific attack exclusively. The majority of the works focus on the bit flips in model parameters \cite{wang2023aegis,li2020defending,he2020defending,rakin2021ra,luo2024deepshuffle,gongye2023hammerdodger,liu2022generating,li2021radar,javaheripi2022acchashtag,javaheripi2021hashtag,hosseini2021safeguarding,liu2023neuropots,liu2020concurrent}. To our best knowledge, there is only one work targeting the DNN executable-level threat~\cite{chen2bitshield}, and no existing defense against bit flips in the computation libraries. Moreover, since the method in~\cite{chen2bitshield} relies on gradient-based techniques, it is ineffective for quantized models where gradients are not computable. A comprehensive solution that can cover bit-flips at different levels of the entire DNN system is urgently needed.

\subsection{Our Contributions}

Driven by the above limitations, we propose \NAME, a novel and holistic methodology to mitigate different levels of BFAs in an efficient and practical manner. The key insight behind \NAME is based on our observation that all BFAs rely on precise identification of vulnerable bits, which constitute a very small percentage (less than 0.01\%) of the total bits (Section \ref{subsec:design_insight}). In contrast, flipping random bits can rarely impact the model inference \cite{gongye2023hammerdodger}. Inspired by this, \NAME aims to \textbf{degrade any carefully-orchestrated BFA into random flips, rendering the attack ineffective}. This is achieved by introducing obfuscation (i.e., random dummy operations) into the application architecture at different levels, including the model parameters, executables, and deep learning libraries. Such obfuscation has minimal impact on the inference results or speed, but alters the memory address offsets of vulnerable bits, thereby invalidating the attacker’s ability to target specific bits. Note that model-level obfuscation has been applied to mitigate side-channel attacks~\cite{li2021neurobfuscator, zhou2022obfunas}. We repurpose this concept and extend it to multiple system layers for integrity protection against BFAs. 

\NAME is designed as an end-to-end protection framework to automatically provide comprehensive protection over a given DNN application (Figure \ref{fig:overview}). This framework consists of three key components: (1) \textit{Vulnerability Searcher} scans the target application, and identifies the potential vulnerable bits at different layers that are sensitive to BFAs. (2) \textit{Obfuscation Pattern Generator} produces the randomized memory address offsets for the vulnerable bits. (3) Finally, \textit{Obfuscation Pattern Enforcer} applies the randomized pattern to the model inference execution at runtime. A set of new algorithms are designed, ensuring the seamless integration of these modules with high efficiency and effectiveness. 

In summary, \NAME offers several key advantages compared to existing methods. First, it provides protection against both model-level and code-level BFAs. Second, injecting dummy operations has zero impact on the model prediction accuracy, and minimally affects the inference latency and resource usage, making \NAME a utility-preserving and lightweight solution. Third, \NAME does not require model retraining or access to training data, rendering it more practical for real-world deployment. We conduct extensive experiments to evaluate \NAME against six state-of-the-art BFAs. The results demonstrate that \NAME can effectively mitigate BFAs across all levels with low overhead, providing robust protection without compromising the performance. Besides, \NAME is able to thwart adaptive attackers who are aware of the defense strategy, with the randomized memory layout obfuscation. 

\section{Background and Related Works}
\label{sec:Background}

\subsection{Bit-Flip Attacks against DNN Models}

Bit-Flip Attacks (BFAs) are a class of hardware fault injection attacks that corrupt the memory contents of a target system by flipping bits, often through the Rowhammer technique \cite{kim2014flipping}. In the deep learning domain, attackers can perform BFAs over critical DNN applications to compromise the prediction results. They can normally achieve two types of consequences: (1) \textit{Untargeted BFAs} seek to significantly degrade a model's overall accuracy. (2) \textit{Targeted BFAs} focus on manipulating the prediction outputs for specific inputs (e.g., belonging to one class, or containing an attacker-defined trigger), while maintaining correct predictions for other inputs. For both untargeted and targeted BFAs, attackers rely on some bit-search algorithms to locate the most vulnerable bits in the target application. Once these bits are identified, they are flipped by the Rowhammer technique. 
BFAs can be implemented at two levels of the target DNN system: 
\begin{itemize}[leftmargin=*,topsep=1pt, itemsep=2pt, itemindent=8pt]
\item\textbf{Model-level Attack.}
This is the most common bit-flip attack strategy against DNN applications. The attackers identify vulnerable bits \textit{within the model parameters} \cite{chen2021proflip,rakin2019bit,rakin2020tbt,yao2020deephammer,rakin2021t,bai2021targeted} to compromise the model behaviors. There are numerous ways for the attacker to look for such bits and associate them with the desired objectives. For instance, TBT~\cite{rakin2020tbt} selects the critical neurons in the final layer of the model, which have the most significant impact on the prediction results. Then it crafts a trigger to activate the selected neurons. Flipping these bits is equivelent to embedding a backdoor into the model. ProFlip \cite{chen2021proflip} searches for bits in all the model layers for flip using the Jacobian Saliency Map Attack (JSMA) technique. TA-LBF \cite{bai2021targeted} formulates the attack as a binary integer programming problem, and adopts the Alternating Direction Method of Multipliers (ADMM) technique to identify the critical bits. 
 
\item\textbf{Code-level Attacks.}
These emerging attacks exploit vulnerable bits in the \textit{binary code} of the DNN application, including the executables generated by deep learning compilers (e.g., TVM \cite{chen2018tvm}, Glow \cite{rotem2018glow}), and the underlying computation libraries. Generally, to support the inference execution of a DNN model, a compiler transforms it from high-level deep learning frameworks (e.g., TensorFlow, PyTorch, ONNX) into machine code targeting specific hardware (e.g., LLVM or NVIDIA CUDA Compiler Driver). The compiled DNN executable may also call some functions from some common libraries for basic computation (e.g., Generalized Matrix Multiply (GEMM) implemented in Basic Linear Algebra Subprograms (BLAS) libraries). By flipping some vulnerable bits in the executables or code libraries, the attackers could hijack the control flow of the target model to perform the wrong computations. 
For instance, Chen et al.\cite{chen2023unveiling} designed an untargeted BFA against DNN executables. It identifies vulnerable bits in the \texttt{.text} section, that could substantially degrade the model performance when flipped. Li et al \cite{li2024yes} designed \textit{FrameFlip} against the dynamic shared libraries like OpenBLAS. Attackers search for vulnerable jump opcodes. By just flipping one single bit in the memory, it can corrupt the control flows of all DNN applications running on top of this library. 
\end{itemize}

\subsection{Existing Defenses against BFAs}  
\label{sec:background-defense}

\subsubsection{Rowhammer Mitigation}
A straightforward solution is to fundamentally fix the hardware vulnerabilities and prevent bit flips in a general way. In particular, multiple solutions have been proposed to prevent the Rowhammer attack. 
One common defense is Error-Correcting Code (ECC), which adds redundancy to stored data for the detection and correction of single-bit errors and, in some cases, double-bit errors. 
An alternative solution is Target Row Refresh (TRR), which periodically refreshes memory rows adjacent to heavily accessed rows, reducing the likelihood of bit flips in neighboring rows. However, both of these methods can be bypassed by the sophisticated many-sided Rowhammer attack~\cite{frigo2020trrespass}. This drives the designs of more advanced defense solutions.
 MASCAT~\cite{irazoqui2016mascat} analyzes the program attributes, e.g., certain instructions or counters, to detect Rowhammer attacks. 
Siloz~\cite{loughlin2023siloz} uses a hypervisor that leverages subarray groups as DRAM isolation domains. 
ZebRAM~\cite{konoth2018zebram} isolates every DRAM row containing data with guard rows that absorb any Rowhammer-induced bit flips. 
Joardar et al.~\cite{joardar2022machine} proposed a machine learning-based approach to analyze counter data and detect attacks. 
Woo et al.~\cite{woo2023scalable} proposed to swap and unswap DRAM rows to mitigate Rowhammer.

Despite their potential, these solutions often require costly hardware upgrades, bring compatibility issues with existing systems and increase complexity. Furthermore, they are only effective for specific devices, but fail to generalize to other computing platforms.

\subsubsection{Defenses Dedicated to BFAs}
To achieve the high generalization and platform-independency, many efforts have been devoted to developing defense solutions dedicated to BFAs. Through our literature survey, we find the majority of existing studies focus on the model-level BFAs exclusively. They can be classified into the following two categories. 

The first strategy is to improve the model robustness against BFAs by revising the DNN architecture or weights.  For instance, 
Aegis~\cite{wang2023aegis} employs a dynamic multi-exit architecture and performs random early exits to reduce the bit-flip impacts on final predictions. 
Li et al.~\cite{li2020defending} proposed a BFA-aware weight reconstruction method that minimizes or diffuses weight perturbations caused by BFAs.
BIN~\cite{he2020defending} adopts binarization-aware training, converting part of the model parameters from high precision to a binary format.
RA-BNN~\cite{rakin2021ra} quantizes the output of activation functions to \{-1, +1\}.
A number of approaches employ the shuffle-based idea to disrupt BFAs.
DeepShuffle~\cite{luo2024deepshuffle} shuffles the channel order in convolutional, linear, and batch normalization layers. HammerDodger~\cite{gongye2023hammerdodger} implements a channel-shuffling mechanism combined with a monitoring system to detect ongoing attacks.
RREC~\cite{liu2022generating} introduces randomized rotations to shuffle the bit order of model weights. 

The second strategy performs error detection and correction with runtime integrity verification. 
RADAR~\cite{li2021radar} derives a 2-bit checksum-based signature for each group of weights within a layer for verification. 
LIMA~\cite{hosseini2021safeguarding} protects the integrity of the most significant bits (MSBs) of DNN weights by embedding integrity marks. 
DeepAttest~\cite{chen2019deepattest} embeds a fingerprint in the DNN weights, which is later verified with a clean one from a Trusted Execution Environment (TEE). 
NeuroPots~\cite{liu2023neuropots} introduces "honey neurons", i.e., crafted vulnerabilities that attract bit flips, paired with a checksum-based detection method to recover model weights from a clean copy stored in the TEE. 
ACCHashtag~\cite{javaheripi2022acchashtag} generates per-layer Pearson hashes of DNN parameters for the top-$k$ most vulnerable layers prior to deployment and verifies them at runtime. 
Hashtag~\cite{javaheripi2021hashtag} encodes weights from vulnerable layers with a low-collision hash function for runtime verification. 
Liu et al.~\cite{liu2020concurrent} proposed a weight encoding-based framework that leverages the spatial locality of bit-flips and encodes sensitive weights for fast detection of BFAs.

In contrast, defense against code-level BFAs is significantly underexplored. To the best of our knowledge, BitShield~\cite{chen2bitshield} is the only existing approach to protecting DNN executables. It employs a Semantic Integrity Guard (SIG) structure to monitor runtime gradients and halt execution if semantics deviate from a predefined range. It also integrates a checksum-based self-Defense mechanism to verify SIG integrity. Currently, there are no solutions to effectively protect the DNN computational libraries against BFAs. 

As discussed in Section \ref{sec:exist-study}, all the above methods suffer from different limitations. These include the impractical requirements of altering the model designs or accessing training data, large impact on the runtime overhead and model performance. More severely, a holistic approach that can cover all types of BFAs are required but missing. This is main objective of our paper.

\section{Overview}

\subsection{Threat Model}
\label{sec:Threat Model}
In line with prior works \cite{liu2023neuropots,yao2020deephammer,rakin2019bit,li2024yes}, we adopt the conventional threat model of BFAs. The attacker is an unprivileged user who shares the same physical machine as the victim’s DNN application. Through techniques like Rowhammer, the attacker can flip the victim's data and code in the DRAM.

We consider a strong attacker, who has common knowledge about the victim's DNN application. (1) As the victim may employ some popular open-source DNN models, the attacker knows the DNN architecture, model weights, gradients and training data. This allows him to precisely target the flip bits within the original DNN model. (2) The victim may also use common deep learning frameworks to support the model inference. Therefore, the attacker knows the framework, dynamic libraries linked to the model, deep learning compilers and compilation settings. (3) The victim employs our \NAME to transform the original DNN model and code executables. The attacker knows the detailed mechanism of \NAME, but not the random numbers generated on-the-fly by the victim. The attacker also does not have access to the transformed DNN model or code executables.

\subsection{Design Insight}
\label{subsec:design_insight}
The design of our \NAME is grounded in the following observation: \textbf{random bit-flips have minimal impact on the model inference process}, which is universal for different levels of the DNN system.  
Specifically, Cheng et al. \cite{gongye2023hammerdodger} showed that randomly flipping bits in DNN's model weights have negligible effects on its prediction results, even when more of bits are changed. We extend this conclusion to the code level by flipping each of all conditional jumps to its opposite control flow in a compiled OpenBLAS library. Out of 154,554 conditional jumps under the \texttt{-O2} optimization flag, only 85 jumps cause a drop in model accuracy when running ResNet-50 on the ImageNet dataset, while another 148 jumps lead to crashes or timeouts. The remaining 99.8\% of the jumps, despite being flipped, run without errors or performance drops. These results suggest that both DNN model weights and codebases are naturally resilient to random bit-flips in most cases.

This inspires the design of \NAME: by introducing obfuscation into the model's inference (i.e., randomizing the memory address offsets for the vulnerable bits), we can transform an organized BFA into random bit-flips, which can effectively alleviate the attacker's impact on the model results. Following this unified principle, \NAME delivers protection over DNN applications against BFAs at both model and code levels.

\subsection{Discussion}
It is worth noting that \NAME fundamentally differs from existing mechanisms involving obfuscation. First, Address Space Layout Randomization (ASLR) \cite{snow2013just} obfuscates the system's virtual memory layout whenever the code execution starts. However, the application binary is still fixed. Therefore, researchers have demonstrated that an attacker can still reliably flip the desired bits in any physical memory page under ASLR based on his knowledge of the binary structure \cite{razavi2016flip, adiletta2024leapfrog}. In contrast, \NAME hardens the \textit{binary representations} to increase the difficulty of targeting vulnerable bit locations. By properly obfuscating the binary, it ensures that the attacker has no knowledge of the vulnerable bit locations within the binary, preventing him from accurately flipping the critical bits. 

Second, software diversity \cite{larsen2014sok,homescu2013profile} aims to create multiple, independent implementations of the same software to enhance security, fault tolerance, and resilience against vulnerabilities. However, it is not effective in mitigating BFAs. (1) Certain software diversity methods (e.g., instruction reordering, equivalent instruction substitution) may not effectively alter memory offsets, leaving the critical bits still vulnerable. (2) DNN computation libraries are highly optimized to minimize overhead in computation-intensive applications. Applying software diversity techniques indiscriminately can disrupt these optimizations, leading to significant performance penalties. (3) Software diversity lacks sensitivity analysis for critical bits and needs to apply obfuscation broadly across the entire software project, which could incur tremendous overhead.

\section{Methodology}
\label{sec:Methodology}

\subsection{Overall Workflow}
\label{subsec:workflow}
\begin{figure}[t]
    \centering
    \includegraphics[scale=0.32]{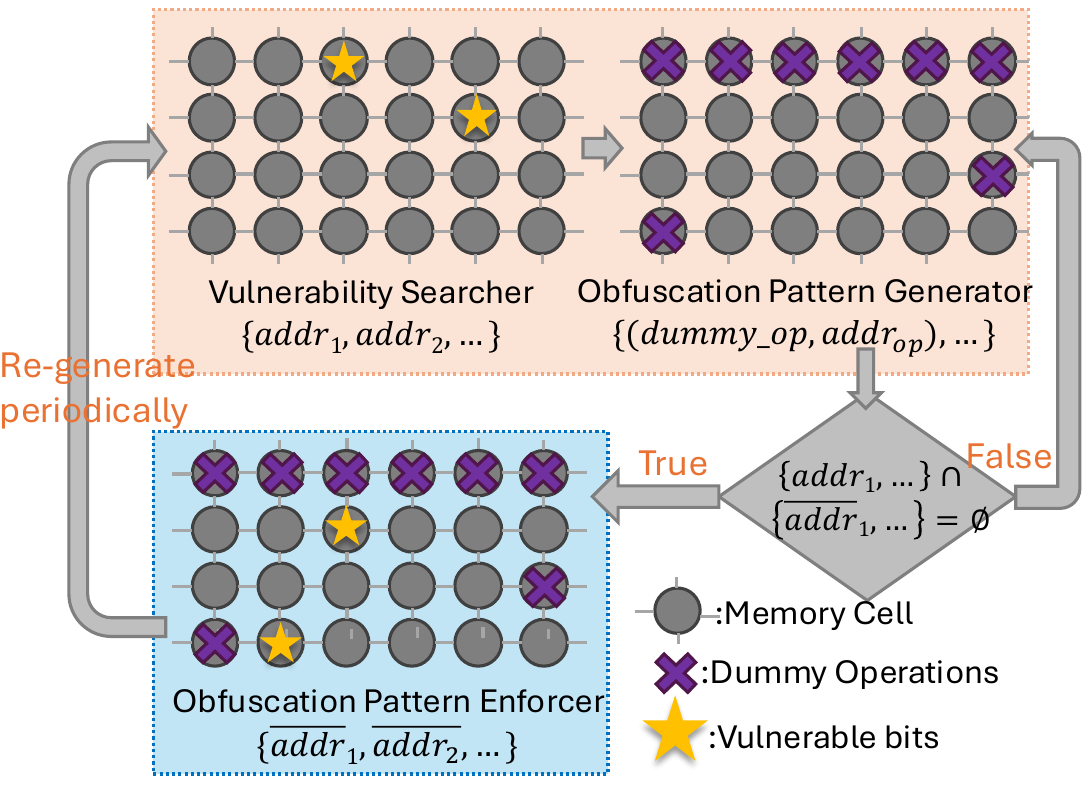}
    \caption{Pipeline of \NAME.}
    \label{fig:overview}
\end{figure}

Figure \ref{fig:overview} shows the pipeline of \NAME. It consists of three logical components: \textit{Vulnerability Searcher}, \textit{Obfuscation Pattern Generator}, and \textit{Obfuscation Pattern Enforcer}. 

\begin{itemize}[leftmargin=*,topsep=1pt, itemsep=2pt, itemindent=8pt]
\item\textbf{Vulnerability Searcher}. This module scans the target application, and records the locations of all identified vulnerable bits. For different levels of BFAs, we design the corresponding algorithms to scan the target operation set (i.e., model weights for model-level BFAs, binaries for code-level BFAs) and identify bits that can significantly impact the model inference when flipped. Vulnerable bits that only cause a crash are not considered critical. The locations (offsets) of these vulnerable bits are recorded in a list.

\item\textbf{Obfuscation Pattern Generator}. Based on these vulnerable bit locations, this module aims to produce the randomized memory address offsets of the target application. The detailed process is outlined in Algorithm \ref{algo:Obfuscation}. Specifically, after \textit{Vulnerability Searcher} obtains the addresses of all vulnerable bits (Line 3), \textit{Obfuscation Pattern Generator} processes the target system and inserts a random number of dummy operations before the critical elements (e.g., operations in the code, layers in the model architecture).

\begin{algorithm}[H]
\small
\caption{Obfuscation Pattern Generation}
\label{algo:Obfuscation}
\begin{algorithmic}[1]
\State \textbf{Input:} \textit{ElemSet}, \textit{Prob} \Comment{Target set, Obfuscation Probability}
\State \textbf{Output:} \textit{ObsPat} \Comment{Obfuscation pattern}
\State ObsSuccess $\gets 0$, VLoc $\gets$ getVulAddr(ElemSet)
\While {ObsSuccess = 0}
    \For {Elem $\in$ ElemSet}
        \If {getRand $\leq$ Prob \textbf{ or } Elem $\in$ VLoc}
            \State (\textit{DummyOp}, \textit{InsLoc}) $\gets$ InsDummy(Op)
            \State ObsPat.append(DummyOp, InsLoc)
        \EndIf
    \EndFor
    \State VLocNew $\gets$ getVulAddr(ObsPat, ElemSet)
    \If {NoCommonAddr between VLoc and VLocNew}
        \State ObsSuccess $\gets 1$
    \EndIf
\EndWhile
\end{algorithmic}
\end{algorithm}

\quad There are two points that need to be emphasized. First, an adaptive attacker may try to flip the bits not in the vulnerable list, attempting to bypass our defense. Such attack will be much less effective as those bits have minor impact on the model behaviors. Nevertheless, we also offer protection for all the other bits not in the list. Note that we cannot enforce the same degree of obfuscation as for the critical bits, which can terribly affect the model performance. Instead, we insert dummy operations following a preset probability, which is typically lower to reduce the overhead. This random insertion mechanism can help further reduce the effectiveness of potential attacks targeting the bits not classified as vulnerable, making them impractical. 

\quad Second, for the generated obfuscation pattern, we run \textit{Vulnerable Searcher} again to check the new vulnerable bits (Line 9). If there is no overlap between the new and original lists of vulnerable bit addresses, the obfuscation is considered successful. This ensures that no vulnerable bits have relocated to new addresses that are also vulnerable. Otherwise, we run \textit{Obfuscation Pattern Generator} again until the condition is satisfied. 

\item\textbf{Obfuscation Pattern Enforcer.}
At runtime, this module applies the generated obfuscation pattern to the model inference. Due to the insertion of dummy operations, the obfuscated operation set will have a different memory layout from the unprotected one. However, this will not affect the functionality of the model, as the inserted dummy operations have no impact on the target operation set. With the randomized memory addresses for critical operations, the attacker cannot identify the actual memory offsets of the vulnerable bits, effectively transforming any sophisticated BFA into an ineffective random bit-flip attack. 

\end{itemize}

\noindent\textbf{End-to-end Protection.}
Given a DNN system, we execute the above three components to generate the obfuscated versions of the model architecture, executable and libraries, and then launch them for robust inference. Note that a sophisticated attacker may try to reverse engineer the obfuscated code, and then adaptively identify new vulnerable bits for flipping. To mitigate this threat, we can periodically run the entire pipeline to generate and launch new obfuscated executables and libraries, e.g., every 10 minutes. Considering the long time for the attacker to flip bits (e.g., multiple hours with Rowhammer), this update frequency is sufficiently secure to mitigate the adaptive BFAs. As our defense pipeline is computationally lightweight and typically completes within a few seconds, the update has negligible impact on the overall performance. We can further launch multiple inference instances, and alternatively update them to reduce the impact on service availability.

Below we present the detailed mechanisms of each component for different levels of BFAs.

\subsection{Vulnerability Searcher}

We design new algorithms to search vulnerable bits in different levels of BFAs, respectively. 

\subsubsection{Model-level BFAs.}

We employ a gradient-based ranking algorithm to identify a set of vulnerable bits in the model weights. For the weights in the \textit{i}-th layer, given the loss function $\mathcal{L}$ and the weight matrix $\mathbf{W}$, the gradient matrix is computed as follows:
\begin{equation}
    \mathbf{G_i}= \frac{\partial \mathcal{L}}{\partial \mathbf{W}} =
\begin{pmatrix}
g_{i}^{1,1} & g_{i}^{1,2} & \dots & g_{i}^{1,N} \\
\vdots & \vdots & \vdots &\vdots \\
g_{i}^{M,1} & g_{i}^{M,2} &  \dots & g_{i}^{M,N} \\
\end{pmatrix}
\end{equation}
We then select the top-\textit{k} weights from both the convolutional layers and the linear layers based on the magnitude of their gradients for the testing dataset. For convolution layers and linear layers from 1 to \textit{z} in the model, the process of selecting the top-\textit{k} most vulnerable weights can be expressed as:
\begin{equation}
\text{Top}_{k}\left[||g_{1}^{1,1}, g_{i}^{1,1}, \dots, g_{z}^{M,N}||\right]
 \end{equation}
The locations of these top-\textit{k} weights form the list of vulnerable locations. Since DNN models are loaded into memory in a sequential page-by-page manner, where each page has a fixed size and stores the weights contiguously, we can map the vulnerable weights directly to their index in memory from the beginning of the model. This indexing allows us to efficiently reference and protect specific weights during our defense implementation.

\subsubsection{Code-level BFAs.}
We aim to search for the critical jump instructions in executables and shared libraries. The process is illustrated in Algorithm \ref{algo:codevulsearch}. 
We focus on jump opcodes as they are highly susceptible to BFAs and flipping a single bit can invert their semantic condition (e.g., changing jump-if-equal to jump-if-not-equal), effectively reversing the control flow. Hence, they are the main target of existing code-based attack~\cite{li2024yes}. The existence of other vulnerable instructions and their sensitivity to bit flips is not explored. To address this uncertainty, \NAME also obfuscates the addresses of non-critical elements not in the vulnerable list, as described in Section~\ref{subsec:workflow}.

\begin{algorithm}[H]
\small
\caption{Binary Automatic Vulnerability Search}
\label{algo:codevulsearch}
\begin{algorithmic}[1]
\State \textbf{Input:} \textit{Code}, \textit{Model}, \textit{Dataset} \Comment{Dynamic library, DL inference infrastructure, Dataset}
\State \textbf{Output:} \textit{v\_loc} \Comment{List of vulnerable bit locations}
\State \textit{v\_loc} $\gets \{\}$, $acc_{clean} \gets \textit{getCleanAcc()}$
\ForAll{functional unit $F \in \textit{Code}$}
    \ForAll{Opcode $Op \in F$}
        \If{$Op \in CondJmpSet$}
            \State $\overline{Op} \gets \textit{flipOpcode}(Op)$, \textit{Instrument}($\overline{Op}$)
            \State $\textit{lib} \gets \textit{writeLibrary()}$, \textit{LibraryInterpositioning}(\textit{lib})
            \State $acc \gets \textit{inference}(model, dataset)$
            \If{$acc \leq acc_{clean}$}
                \State \textit{v\_loc} $\gets \textit{v\_loc} \cup \{\textit{v\_loc}\}$
            \EndIf
            \State \textit{recover}($Op$)
        \EndIf
    \EndFor
\EndFor
\State \Return \textit{v\_loc}
\end{algorithmic}
\end{algorithm}

Specifically, we iterate through all opcodes in the binary files. If the current opcode is a conditional jump, we flip the conditional jump to its semantic opposite and generate a new library file. In the x86 instruction set, each conditional jump has a short jump and near jump version, which have different ranges and opcodes. For example, \texttt{je} has the short jump opcode \texttt{0x74} and the near jump opcode \texttt{0x0F84}. The algorithm considers both cases. We then use the instrumented binary to run inference and test accuracy. If the accuracy drops, the branch is considered vulnerable, and the corresponding bit address is recorded.

\subsection{Obfuscation Pattern Generator}
We develop new obfuscation solutions to randomize the memory offsets of the identified critical operations. 
\subsubsection{Model-level BFAs.}
We introduce two strategies to obfuscate the model address layouts, as shown in Figure \ref{fig:modelobfuscation}.

\begin{itemize}[leftmargin=*,topsep=1pt, itemsep=2pt, itemindent=8pt]
\item\textbf{Inserting Dummy Layers.}
This strategy involves inserting an additional computational layer immediately after the activation function \(\varphi\) of a given layer. The insertion of a dummy layer \(L_i\) does not affect the output of the original layer, as shown in Equation \ref{eq:dummylayer}:
\begin{equation}
    \label{eq:dummylayer}
    \varphi( \varphi(X_i \cdot W_i) \cdot L_i) = \varphi(X_i \cdot W_i)
\end{equation}

\quad For linear layers, the dummy layer \(L_i\) is an identity matrix of the same dimensions as the input to the layer. For convolutional layers, where the input and output have \textit{c} channels and kernel size of \((k_1, k_2)\), the dummy layer is initialized as:
\[
L_{i}^{(a,b,d,m)} = 
\begin{cases}
1 & \text{if } a = b = 1 \text{ and } d = m \\
0 & \text{otherwise}
\end{cases}
\]
where \(a, b\) are the indices of the row and column, \(d\) is the index of the input channel, and \(m\) is the index of the filter. We choose a kernel size of \(k_1 = k_2 = 1\), which minimizes the extra computation introduced by the dummy layer.

\begin{figure}[t]
    \centering
    \includegraphics[scale=0.068]{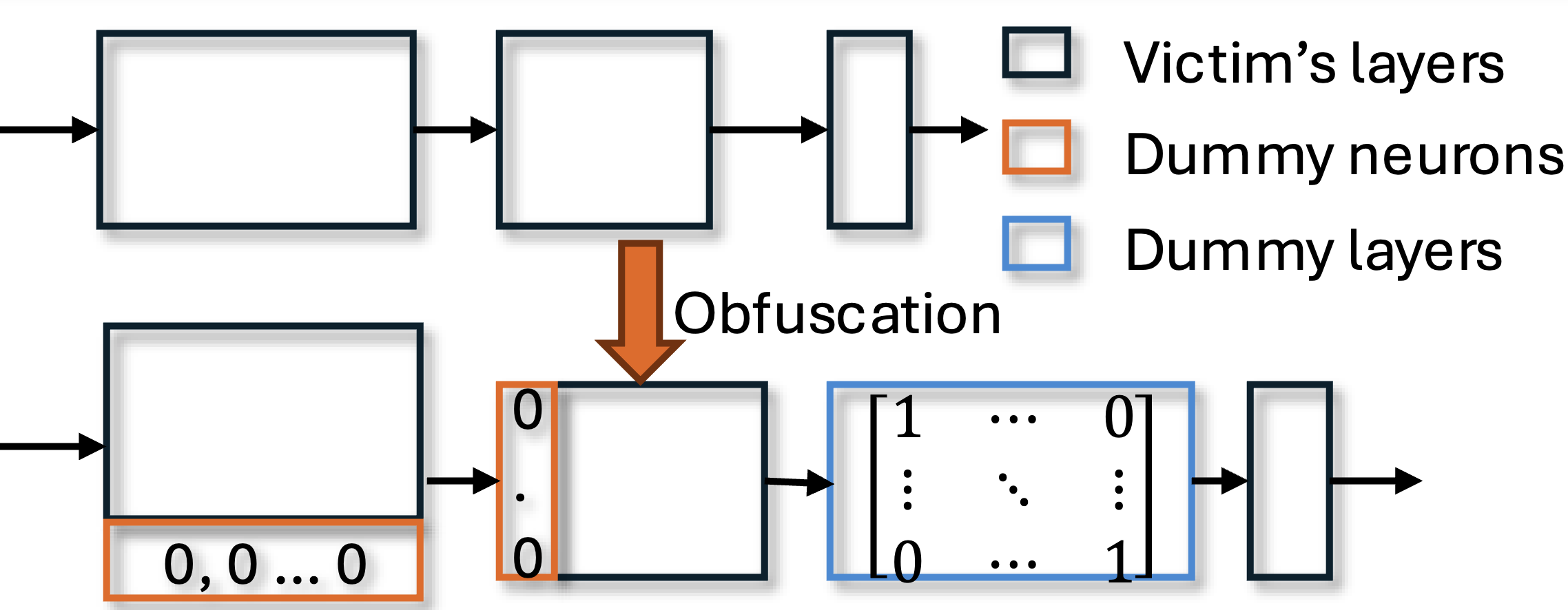}
    \caption{Model Obfuscation.}
    \label{fig:modelobfuscation}
\end{figure}

\item\textbf{Inserting Dummy Neurons.}
Another obfuscation strategy is the insertion of dummy neurons into existing layers. These dummy neurons ensure that the prediction behavior of the original DNN remains unchanged after their addition.

\end{itemize}

To insert \textit{n} dummy neurons into a current computation layer with weights \(W_i^{(b,a)}\), assuming the next computation layer has weights \(W_{i+1}^{(c,b)}\), then given the input \(X_i\), the forward process is formulated as:
\begin{equation}
    \label{eq:dummyneuron}
    W_{i+1}^{c,b} \cdot (W_{i}^{b,a} \cdot \mathbf{x}) = W_{i+1}^{c,(b+n)} \cdot (W_i^{(b+n),a} \cdot \mathbf{x}) 
\end{equation}
To ensure the injected neurons are non-functional (i.e., "dummy"), their weights are set to vectors of all zeros with biases of zero. Since adding dummy neurons alters the dimensions of the current layer, the subsequent layer must also have the corresponding dummy neurons added to maintain the dimensional consistency. 
Specifically, for a convolution layer, we add \textit{n} extra filters to the current layer, increasing the number of output channels by \textit{n}. As a result, the number of input channels of the next convolution layer must also be increased by \textit{n}, with the additional channels initialized to zeros.
Similarly, for each linear layer, we increase the number of output features by \textit{n}, and adjust the number of input features of the next linear layer by \textit{n}, with the added parameters initialized to zeros.

\subsubsection{Code-level BFAs.}

To add obfuscation at the code level, our strategy is to insert dummy operations (e.g., \textit{DummyOP}) before the critical instructions. The \texttt{NOP} (No Operation) instruction in \texttt{x86} architecture is a one-byte instruction that takes up space in the instruction stream but performs no useful operation. It does not modify any machine state except for advancing the \texttt{EIP} (Extended Instruction Pointer) register. As a result, after we determine that a specific location requires obfuscation, a random number of \texttt{NOP} instructions are inserted before the current opcode, thereby altering the memory layout without changing the functionality.

\begin{figure}[t]
    \centering
    \includegraphics[scale=0.42]{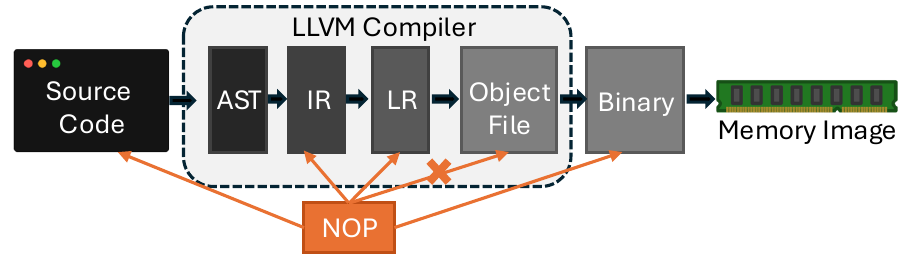}
    \caption{Possible NOP insertion phases during compiling.}
    \label{fig:llvm}
\end{figure}

To implement this memory address randomization efficiently, we develop an LLVM-based backend obfuscation pass. LLVM is a suite of compiler and toolchain technologies that transform source code (e.g., C programs) into an  Abstract Syntax Tree (AST), which is a structured representation. Then it is translated into intermediate representation (IR), which is independent of the source language. This IR is optimized by LLVM passes, after which it is further lowered into a lower-level representation (LR), such as GCC’s Register Transfer Language (RTL). The backend then generates object files from this optimized IR. Finally, a linker combines object files into the final program binary, which the operating system's loader places in memory for execution. Figure \ref{fig:llvm} illustrates the compilation stages of a source file, highlighting the possible phases for \texttt{NOP} insertion. 

While it is theoretically possible to insert \texttt{NOP} instructions at any stage of the compilation process, it is challenging to do so in earlier stages (e.g., at the IR level or in source code). Since the attack targets the memory image, corresponding to the final binary, it is important to ensure precise control over the location of inserted \texttt{NOP} instructions in the binary. However, locating the exact binary position for vulnerability obfuscation from the IR level is difficult due to the abstraction gap between the IR and the final binary. Inserting \texttt{NOP} instructions at the source code level also presents challenges. Even if the mapping between the source code and binary is known, compiler optimizations can relocate or even remove the inserted code, making it difficult to control where the \texttt{NOP} instructions will eventually appear in the binary. 

To overcome these challenges, we insert \texttt{NOP} instructions at the lower-level representation, specifically after the compiler performs all optimizations and just before the target code is emitted (as indicated by "x" in Figure \ref{fig:llvm}). This step corresponds to the \texttt{addPreEmitPass2} in LLVM, ensuring that the inserted \texttt{NOP} instructions appear precisely where required in the final binary. To manage the \texttt{NOP} insertion process, we store the insertion patterns in a file, which contains both the insertion locations and the number of \texttt{NOP} instructions to be inserted. This file is then loaded by our backend pass during compilation to effectively obfuscate the memory layout. 

\subsection{Obfuscation Pattern Enforcer}

\subsubsection{Model-level BFAs}
At runtime, we load the target model and apply the obfuscation patterns to disrupt the memory layout. Since the obfuscation involves only simple operations, e.g., padding zeros to the weights or inserting non-functional neurons, it is computationally efficient and incurs minimal overhead. As a result, the transformation is applied almost instantaneously, leaving very little time for an attacker to exploit the model via Rowhammer or other bit-flip techniques before the obfuscation is completed.

Moreover, due to the design of the obfuscation strategies, the obfuscated model maintains identical functionality to the original model. This ensures that the obfuscation process does not introduce any unintended changes to the model's behavior, preserving its accuracy and performance while simultaneously protecting it against memory-based attacks. Additionally, the memory layout of the model is randomized with each load, making it even more challenging for an attacker to predict the specific bit locations they need to target. This randomness further strengthens the defense.

\subsubsection{Code-level BFAs}
Once the instrumented library or executable with the altered memory layout is generated, we load them for runtime model inference. In Linux environments, when a program is loaded and executed, the dynamic linker (ld-linux.so) first searches the libraries listed in the \texttt{LD\_PRELOAD} environment variable for any undefined references before searching other libraries. Therefore, we set \texttt{LD\_PRELOAD} to our obfuscated versions, which can intercept and redirect function calls to utilize the new libraries. For code executables, we modify the compilation process, e.g., instruct TVM to generate LLVM bitcode, enabling our custom LLVM backend pass to apply the obfuscation techniques. 

We can frequently regenerate libraries and executables to enhance protection. The compilation cost can be very small with the following optimization. Since the functions prone to BFAs are concentrated in a small subset of source files (8 in OpenBLAS), we only need to recompile these files instead of the entire project. To further reduce the compilation time, we pre-generate LLVM intermediate representation (IR) and perform binary generation from this IR, as the obfuscation is applied in the backend, specifically in the IR-to-binary translation process. Our tests show that the compilation process takes around 0.683 seconds, and generating the dynamic library adds just 0.103 seconds. This enables frequent, low-cost binary updates, maximizing the runtime protection against BFAs.

\section{Security Analysis}
\label{sec:analysis}

In this section, we provide analysis about the security robustness of \NAME from different perspectives. 

\begin{itemize}[leftmargin=*,topsep=1pt, itemsep=2pt, itemindent=8pt]
\item\textbf{Secrecy of obfuscation}. We assume that our detailed defense mechanism (e.g., algorithms, type of obfuscation primitives) is totally public and known by attackers. However, they cannot predict the exact locations or quantities of obfuscation patterns, which are randomly generated on-the-fly. This randomness in our design can ensure its security. This is analogous to the Kerckhoffs's principle in cryptographic systems, where knowing the algorithm does not compromise the security without the key. 

\item\textbf{Reverse-engineering the application}. 
Even the attacker has no knowledge or white-box access to the obfuscated DNN application, he may adopt some reverse-engineering techniques to recover it \cite{hua2018reverse}, further finding the hidden vulnerable bits and then flipping them. As described in Section \ref{subsec:workflow}, we can easily block this attack path by periodically updating the application with new obfuscation patterns. Note that each update only takes less than one second, while the attacker may spend dozens of seconds to flip the bits with RowHammer \cite{yao2020deephammer}. Such imbalanced timing cost makes this adaptive attack impractical. 

\item\textbf{Targeting a different set of bits}.
Theoretically, our \textit{Vulnerability Searcher} can identify majority of vulnerable bits, and attacker’s identified bits highly likely fall within this set. Even if the attacker selects a different set of bits to bypass the defense (the corresponding attack will be much less effective), our design (Algorithm \ref{algo:Obfuscation}) ensures obfuscation is applied probabilistically throughout the entire codebase or model weights. It will largely alter the memory offsets to thwart the attack. 

\item\textbf{Flip bits of dummy neurons}.
It is possible that attackers may flip bits of the inserted dummy neurons, which could affect the model behaviors. However, this situation is very rare as the inserted dummy weights constitute a negligible fraction of the total weights. In our evaluations, this never occurred. To further mitigate this risk, we can refine the obfuscation algorithm to exert finer control over insertion locations, minimizing the likelihood of such occurrences.

\item\textbf{Triggering side-channel attacks.}
The obfuscated models may introduce GPU alignment issues, potentially increasing computation time. This might lead to side-channel leakage. Although side-channel attack is beyond the scope of this paper, we still believe such threat is infeasible to exploit in practice, due to the following reasons. (1) The attackers do not have the timing profile of the victim's original model. Without such baseline, it is difficult for them to distinguish the long execution time is attributed to alignment issues, or simply normal larger layer. (2) Even if the attackers know a layer is obfuscated, the randomness introduced by \NAME substantially expands the search space, prevents them from inferring the structure of the original layer. (3) In fact, model obfuscation has been successfully employed in prior works \cite{li2021neurobfuscator,zhou2022obfunas}, as effective countermeasures against side-channel-based model extraction attacks. This further supports that the obfuscation of \NAME will not enlarge the attack surface. 

\item\textbf{Generalization to other ISAs.}
Although our implementation for the code-level obfuscation mainly focuses on the x86 ISA, \NAME can be easily extended to other ISAs. First, the \texttt{NOP} operation is universally available across most ISAs for memory alignment and timing purposes. Second, since \texttt{NOP} insertion is performed via an LLVM backend pass, and LLVM supports a wide range of ISAs, adapting \NAME to another ISA requires minimal modifications.

\end{itemize}

\section{Evaluation}
\label{sec:Evaluation}

\subsection{Experimental Setup}
\textbf{Implementation.}  
Our experiments were conducted on an Nvidia GeForce RTX 3090 GPU with 24GB of memory, used for deep learning model training. The models were deployed on a server powered by an AMD EPYC 7513 32-Core Processor for inference. The DNN models and our defense are implemented with PyTorch version 1.13 with CUDA 11.6, compiled using LLVM version 12.0.1. 

\noindent\textbf{Attacks under Evaluation.}
For \textit{code-level BFAs}, we assess the effectiveness of \NAME against two state-of-the-art, untargeted attacks, which are the only known attacks of their kind to date. (1) \textit{FrameFlip} \cite{li2024yes} targets the BLAS library by flipping conditional jumps in LLVM bitcode via an \texttt{XOR} operation at the IR level. We used PyTorch version 2.3.0 with OpenBLAS version 0.3.20 as the underlying library for optimized linear algebra computations. To better reflect real-world threats, we tested our defense on a C++-based inference infrastructure, as used in \cite{li2024yes}, to demonstrate its applicability beyond Python-based frameworks. 
(2) The attack in~\cite{chen2023unveiling} targets the TVM compiler by manipulating vulnerable bits in DL executables to degrade overall model accuracy. We adopt the TVM version 0.18.dev0, and choose the \texttt{-O3} optimization flag. We also include adaptive versions of both attacks to evaluate \NAME's robustness under adversarial awareness.

For \textit{model-level BFAs}, we evaluate \NAME against one untargeted attack \cite{rakin2019bit}, and three targeted attacks: T-BFA\cite{rakin2021t}, TBT~\cite{rakin2020tbt}, and TA-LBF~\cite{bai2021targeted}. As with the code-level setting, we also consider adaptive variants of these attacks.

\noindent\textbf{Datasets and Models.}
Our solution is general over different tasks. Following common selections in existing attacks~\cite{li2024yes, chen2023unveiling, chen2021proflip, rakin2019bit, rakin2020tbt, yao2020deephammer, rakin2021t, bai2021targeted} and defenses~\cite{luo2024deepshuffle, liu2023neuropots, gongye2023hammerdodger, liu2022generating, he2020defending, rakin2021ra, wang2023aegis}, we evaluate \NAME on three popular datasets: CIFAR-10 \cite{krizhevsky2009learning}, German Traffic Sign Recognition Benchmark (GTSRB) \cite{houben2013detection}, and ImageNet \cite{russakovsky2015imagenet}. CIFAR-10 consists of 60,000 colored images across 10 classes with size 32×32, including 50,000 images for training and 10,000 images for testing. GTSRB has 43 traffic sign classes with 39,209 training images and 12,630 test images. For ImageNet, we use the ILSVRC-2012 subset, which includes 1,281,167 colored training images and 50,000 evaluation images, each of size 224×224, across 1,000 classes. 

We choose popular network architectures that are widely used for image classification tasks, including VGG-16~\cite{simonyan2014very}, ResNet-20, ResNet-32, ResNet-34, and ResNet50~\cite{he2016deep}.

\noindent\textbf{Baselines.}
For \textit{code-level BFAs}, we compare our \NAME with the only known defense BitShield~\cite{chen2bitshield}. For \textit{model-level BFAs}, we compare \NAME against several state-of-the-art defenses targeting both untargeted and targeted bit-flip attacks on model weights.
Specifically, for untargeted BFAs, we choose DeepShuffle~\cite{luo2024deepshuffle}, NeuroPots~\cite{liu2023neuropots}, HammerDodger~\cite{gongye2023hammerdodger},  and RREC~\cite{liu2022generating}.
For targeted BFAs, we choose BIN~\cite{he2020defending}, RA-BNN~\cite{rakin2021ra}, and Aegis~\cite{wang2023aegis}.

\subsection{Results for Code-level BFAs}
\label{sec:Results for Code-based Bit-flip Attacks}

\noindent\textbf{Model Utility Evaluation.}  
A key metric for evaluating any defense is its ability to preserve the model’s utility, particularly in terms of accuracy (Acc). Tables \ref{tab:eval_code_util} and \ref{tab:eval_exe_util} present the accuracy results after applying \NAME across different DL frameworks (C++-based and PyTorch-based) with the TVM compiler. 
The "Base Acc" column shows the model's accuracy without \NAME, while the "$\Delta$Acc" column represents the change in accuracy after applying \NAME. The "Flag" column indicates the optimization flag used during library compilation. The results demonstrate that \NAME does not degrade model performance, with no observable accuracy loss across all datasets and models tested.

\begin{table}[htbp]
  \centering
  \caption{Model utility evaluation (DL infra,)}
    \resizebox{.85\linewidth}{!}{
    \begin{tabular}{cclrr}
    \toprule
    \multicolumn{1}{l}{DL} & \multirow{2}[2]{*}{Flag} & \multirow{2}[2]{*}{Dataset-Model} & \multicolumn{1}{l}{Base } & \multicolumn{1}{l}{$\Delta$Acc} \\
    \multicolumn{1}{l}{Framework} &  &       & \multicolumn{1}{l}{Acc} & \multicolumn{1}{l}{ (\%)} \\
    \midrule
    \multirow{9}[2]{*}{C++} & \multirow{3}[1]{*}{O1} & CIFAR10-ResNet20 & 97.00    & 0 \\
          &       & GTSRB-VGG16 & 98.24  & 0 \\
          &       & ImageNet-ResNet50 & 91.67  & 0 \\
          & \multirow{3}[0]{*}{O2} & CIFAR10-ResNet20 & 97.00    & 0 \\
          &       & GTSRB-VGG16 & 98.24  & 0 \\
          &       & ImageNet-ResNet50 & 91.67  & 0 \\
          & \multirow{3}[1]{*}{O3} & CIFAR10-ResNet20 & 97.00    & 0 \\
          &       & GTSRB-VGG16 & 98.24  & 0 \\
          &       & ImageNet-ResNet50 & 91.67  & 0 \\
    \midrule
    \multirow{9}[2]{*}{Pytorch} & \multirow{3}[1]{*}{O1} & CIFAR10-ResNet20 & 81.25  & 0 \\
          &       & GTSRB-VGG16 & 93.81  & 0 \\
          &       & ImageNet-ResNet50 & 87.50  & 0 \\
          & \multirow{3}[0]{*}{O2} & CIFAR10-ResNet20 & 81.25  & 0 \\
          &       & GTSRB-VGG16 & 93.81  & 0 \\
          &       & ImageNet-ResNet50 & 87.50  & 0 \\
          & \multirow{3}[1]{*}{O3} & CIFAR10-ResNet20 & 81.25  & 0 \\
          &       & GTSRB-VGG16 & 93.81  & 0 \\
          &       & ImageNet-ResNet50 & 87.50  & 0 \\
    \bottomrule
    \end{tabular}%
    }
  \label{tab:eval_code_util}%
\end{table}%

\begin{table}[htbp]
  \centering
  \caption{Model utility evaluation (DL compiler TVM)}
    \begin{tabular}{lrr}
    \toprule
    Dataset-Model & \multicolumn{1}{l}{Base Acc (\%)} & \multicolumn{1}{l}{$\Delta$Acc (\%)} \\
    \midrule
    CIFAR10-ResNet20 & 80.50 & 0 \\
    GTSRB-VGG16 & 98.50 & 0 \\
    ImageNet-ResNet50 & 85.00 & 0 \\
    \bottomrule
    \end{tabular}%
  \label{tab:eval_exe_util}%
\end{table}%

\noindent\textbf{Mitigating Untargeted Attacks.} 
For untargeted BFAs, we measure the average accuracy drop across the identified vulnerable bit set. Tables \ref{tab:eval_code_mitigate_untarget_c}, \ref{tab:eval_code_mitigate_untarget_pytorch}, and \ref{tab:eval_exe_untarget} report the number of identified vulnerable bits (\#Vuln. Bits), the original model accuracy (before the attack), and the accuracy both without and with the defense applied to the vulnerable bit set. 
These tables demonstrate that \NAME successfully restores the model to its original accuracy for both C++-based and PyTorch-based DL infrastructures under the \textit{FrameFlip} attack, and for the TVM compiler under the attack in \cite{chen2023unveiling}.

\begin{table}[htbp]
  \centering
  \caption{Mitigating untargeted attacks (C++ infra.)}
  \resizebox{.8\linewidth}{!}{
    \begin{tabular}{clrrrr}
    \toprule
    \multirow{2}[4]{*}{Flag} & \multirow{2}[4]{*}{Model} & \multicolumn{1}{l}{\#Vuln} & \multicolumn{3}{c}{Acc (\%)} \\
\cmidrule{4-6}          &       & \multicolumn{1}{l}{Bits} & \multicolumn{1}{l}{Base} & \multicolumn{1}{l}{w/o. Def} & \multicolumn{1}{l}{w. Def} \\
    \midrule
    \multirow{3}[1]{*}{O1} & ResNet20 & 80    & 97.00  & 33.90  & 97.00 \\
          & VGG16    & 61    & 98.24  & 18.00  & 98.24 \\
          & ResNet50 & 81 & 91.67  & 29.10  & 91.67 \\
    \midrule
    \multirow{3}[1]{*}{O2} & ResNet20 & 85    & 97.00  & 36.50  & 97.00 \\
          & VGG16    & 64    & 98.24  & 22.50  & 98.24 \\
          & ResNet50 & 85 & 91.67  & 31.20  & 91.67 \\
    \midrule
    \multirow{3}[1]{*}{O3} & ResNet20 & 85    & 97.00  & 36.20  & 97.00 \\
          & VGG16    & 63    & 98.24  & 23.50  & 98.24 \\
          & ResNet50 & 85 & 91.67  & 31.60  & 91.67 \\
    \bottomrule
    \end{tabular}%
    }
  \label{tab:eval_code_mitigate_untarget_c}%
\end{table}%

\begin{table}[htbp]
  \centering
  \caption{Mitigating untargeted attacks (PyTorch infra.)}
  \resizebox{.8\linewidth}{!}{
    \begin{tabular}{clrrrr}
    \toprule
    \multirow{2}[4]{*}{Flag} & \multirow{2}[4]{*}{Model} & \multicolumn{1}{l}{\#Vuln} & \multicolumn{3}{c}{Acc (\%)} \\
\cmidrule{4-6}          &       & \multicolumn{1}{l}{Bits} & \multicolumn{1}{l}{Base} & \multicolumn{1}{l}{w/o. Def} & \multicolumn{1}{l}{w. Def} \\
    \midrule
    \multirow{3}[1]{*}{O1} & ResNet20 & 34    & 81.25  & 23.67 & 81.25 \\
          & VGG16    & 30    & 95.31  &  6.88 & 95.31 \\
          & ResNet50 & 28 & 87.50  &  5.58 & 87.50 \\
    \midrule
    \multirow{3}[1]{*}{O2} & ResNet20 & 34    & 81.25  & 25.14 & 81.25 \\
          & VGG16    & 34    & 95.31  & 11.81 & 95.31 \\
          & ResNet50 & 29 & 87.50  &  6.25 & 87.50 \\
    \midrule
    \multirow{3}[1]{*}{O3} & ResNet20 & 34    & 81.25  & 25.18 & 81.25 \\
          & VGG16    & 31    & 95.31  &  8.06 & 95.31 \\
          & ResNet50 & 28 & 87.50  &  7.14 & 87.50 \\
    \bottomrule
    \end{tabular}%
    }
  \label{tab:eval_code_mitigate_untarget_pytorch}%
\end{table}%

\begin{table}[htbp]
  \centering
  \caption{Mitigating untargeted attacks (TVM Compiler)}
    \begin{tabular}{lrrrr}
    \toprule
    \multirow{2}[4]{*}{Model} & \multicolumn{1}{l}{\#Vuln.} & \multicolumn{3}{c}{Acc (\%)} \\
\cmidrule{3-5}          & \multicolumn{1}{l}{Bits} & \multicolumn{1}{l}{Base} & \multicolumn{1}{l}{ w/o. Def.} & \multicolumn{1}{l}{w. Def.} \\
    \midrule
    ResNet20 & 94    & 80.50 & 17.98 & 80.50 \\
    VGG16 & 112   & 98.50 & 16.83  & 98.50 \\
    ResNet50 & 72    & 85.00 & 21.25 & 85.00 \\
    \bottomrule
    \end{tabular}%
  \label{tab:eval_exe_untarget}%
\end{table}%

\noindent\textbf{Adaptive Attacks.}  
An effective defense mechanism should withstand adaptive attackers who are aware of its existence and functionality. We consider such an attacker who understands the details of our defense mechanism (but not the obfuscated parameters generated on-the-fly) and aims to develop an adaptive strategy to bypass it. Specifically, he is aware that our defense randomly obfuscates the memory address layout by inserting \texttt{nop} operations.
Since \NAME introduces randomness into the memory layout obfuscation, the attacker’s only viable option to circumvent it is to flip more bits. To this end, he flips not only the bits in the identified vulnerable bit list but also adjacent bits around each vulnerable bit address \textit{addr}.
Given that code alignment is typically enforced by compilers (e.g., aligning instructions to 16-byte boundaries for efficient CPU instruction fetching), \NAME may cause 16\textit{n}-byte latencies in subsequent function. Therefore, the attacker can flip all bits in the following range for each vulnerable bit address \textit{addr}:
\begin{equation}
    [addr - x_1 + x_2 \cdot al, addr + x_1 + x_2 \cdot al]
\end{equation}
where \textit{al} represents the alignment size set during compilation, and \(x_1\) and \(x_2\) are parameters that control the flipping range. The larger value results in a broader range of bit flips.

We conducted such adaptive attack on all three dataset-model configurations using the compile flags \texttt{-O1}, \texttt{-O2}, and \texttt{-O3} on the C++ deep learning framework, as well as on all three dataset-model configurations in the DL compiler TVM. For each vulnerable location \textit{addr}, we set the alignment size \textit{al} to 16, and evaluate \(x_2\) in the range [0, 1, 2] for each evaluation, while \(x_1\) takes values from the set \{5, 10, 15, 20, 25, 30, 35, 40, 45\} at each time. We observed an increase in the percentage of program timeouts or crashes, but no accuracy drop was detected across any experimental setting.

These results demonstrate that even when an attacker is aware of our defense mechanism, he is unable to bypass it. The adoption of randomized memory address obfuscation turns carefully targeted bit flips into random flips, neutralizing the attacker’s adaptive strategy and preserving model accuracy.

\noindent\textbf{Overhead.}
A key design objective of \NAME is to minimize the overhead. We evaluate the time, storage, and memory overhead introduced by \NAME. For the time overhead, we conduct 10 inference runs for each model and compute the average increase in inference time per image. For the storage overhead, we measure the increase in the size of the dynamically shared library files due to the insertion of \texttt{nop} operations. For the memory overhead, we conduct 10 inference runs for each model and compute the average increase in memory consumption per image.

The results are shown in Tables \ref{tab:eval_code_overhead} and \ref{tab:eval_exe_overhead}. The "Insert Count" column represents the number of \texttt{nop} instructions inserted into the dynamically linked libraries. The "$\Delta$Time" column shows the change in inference time per image, "$\Delta$Storage" reflects the increase in file size due to the added \texttt{nop} instructions, and the "$\Delta$Memory" column shows the change in inference memory consumption per image. 
We observe that \NAME introduces minimal overhead in all aspects. The time overhead varies slightly, with some models benefiting from micro-optimizations that reduce inference time, while others, such as ImageNet-ResNet50 under \texttt{-O3}, gets small increase (5.02\%). The storage overhead is negligible in the C++ framework, with only minor file size increases ($\leq$0.37\%), and remains within an acceptable range in the TVM framework. Memory consumption exhibits slight fluctuations across models and optimization levels, with a maximum recorded increase of 1.69\% for VGG16 under \texttt{-O3} in the C++ framework and 2.23\% in the TVM compiler. Overall, these findings confirm that \NAME imposes minimal execution and resource overhead, ensuring its feasibility for deployment without significant impact on system performance.

\begin{table}[htbp]
  \centering
  \caption{Time and storage overhead for C++ infra..}
  \resizebox{\linewidth}{!}{
    \begin{tabular}{lrrrrr}
    \toprule
    \multirow{2}[2]{*}{Flag} & \multirow{2}[2]{*}{Model} & \multicolumn{1}{l}{Insert} & \multicolumn{1}{l}{$\Delta$Time} & $\Delta$Storage & $\Delta$Memory\\
          &       & \multicolumn{1}{l}{Count} & \multicolumn{1}{l}{(ms/\%)} & (KB/\%)  & (KB/\%)\\
    \midrule
    \multirow{3}[1]{*}{O1} & ResNet20 & 63636 & +2(3.04\%)   & +47(0.37\%)& -13(-0.46\%) \\
          & VGG16 & 34664 & -1(-1.08\%)   & +27(0.22\%)& +4(1.00\%) \\
          & ResNet50 & 37085 & +14(1.31\%)    & +27(0.22\%)& +195(1.00\%) \\
    \multirow{3}[0]{*}{O2} & ResNet20 & 55667 & -8(-13.2\%)     & +39(0.29\%)& +30(1.05\%) \\
          & VGG16 & 4939  & -4(-3.13\%)      & +4(0.03\%)& -7(-1.67\%) \\
          & ResNet50 & 33494 & -42(-3.90\%)      & +23(0.17\%)& -1,132(-6.01\%) \\
    \multirow{3}[1]{*}{O3} & ResNet20 & 56393 & -8(-12.8\%)      & +39(0.26\%)& +67(2.36\%) \\
          & VGG16 & 5867  & -2(-1.26\%)       & +4(0.03\%) & +7(1.69\%)\\
          & ResNet50 & 5272  & +51(5.02\%)      & +4(0.03\%)& -153(-0.80\%) \\
    \bottomrule
    \end{tabular}%
    }
  \label{tab:eval_code_overhead}%
\end{table}%

\begin{table}[htbp]
  \centering
  \caption{Time and storage overhead for TVM compiler.}
    \begin{tabular}{lrrrr}
    \toprule
    \multirow{2}[1]{*}{Model} & \multicolumn{1}{l}{Insert} & \multicolumn{1}{l}{$\Delta$Time } & \multicolumn{1}{l}{$\Delta$Storage} & $\Delta$Memory\\
          & \multicolumn{1}{l}{Count} & \multicolumn{1}{l}{(ms/\%)} & \multicolumn{1}{l}{(KB/\%)} & (KB/\%)\\
          \midrule
    ResNet20 & 39404      & +398(2.79\%)      &+31(7.58\%) &+12(0.67\%) \\
    VGG16 & 31248      &  +1720(4.45\%)      &+39(10.69\%) &+173(2.23\%) \\
    ResNet50 & 9989      &  -743(-0.55\%)      &+8(0.69\%) &+6(0.18\%) \\
    \bottomrule
    \end{tabular}%
  \label{tab:eval_exe_overhead}%
\end{table}%

\noindent\textbf{Comparison.} 
We compare \NAME with BitShield~\cite{chen2bitshield}, the only existing defense at the DNN executable level. We use the ResNet50 model on ImageNet for the BFA against the TVM compiler. The comparison results are shown in Table~\ref{tab:eval_exec_untarget_compare}. Both defenses maintain the model’s original accuracy ($\Delta$Acc = 0\%). However, BitShield introduces a notable 8.22\% runtime overhead, while \NAME incurs a slight performance improvement of -0.55\%, possibly due to compiler-level code alignment optimizations introduced by obfuscation. In terms of mitigation effectiveness, \NAME completely thwarts the attack with a 100\% mitigation rate, while BitShield achieves 97.9\%. These highlight that \NAME not only offers superior protection but also improves efficiency in certain cases.

\begin{table}[htbp]
  \centering
  \caption{Comparison with Bitshield}
    \begin{tabular}{llll}
    \toprule
    {Defense} & {$\Delta$Acc(\%)} & {$\Delta$Time(\%)}  & {Mitigation Rate(\%)} \\
    \midrule
    Bitshield  & 0 & 8.22    & 97.9\\
    \NAME & 0 & \textbf{-0.55}    & \textbf{100}\\
    \bottomrule
    \end{tabular}%
  \label{tab:eval_exec_untarget_compare}%
\end{table}%

\subsection{Results for Model-level BFAs}
\label{sec:Results for Model-based Bit-flip Attacks}

\noindent\textbf{Model Utility Evaluation.}  
We applied \NAME to model-level BFAs and evaluated its impact on model functionality, specifically in terms of accuracy. The results, shown in the Table \ref{tab:eval_model_util}, indicate that \NAME barely degrades the performance of the original model.
\begin{table}[htbp]
  \centering
  \caption{Model utility evaluation after applying \NAME.}
    \begin{tabular}{lrr}
    \toprule
    Dataset-Model & \multicolumn{1}{l}{Base Acc (\%)} & \multicolumn{1}{l}{$\Delta$Acc (\%)} \\
    \midrule
    CIFAR10-ResNet20 & 85.36 & +0.01 \\
    CIFAR10-ResNet32 & 84.38 & +0.01 \\
    CIFAR10-ResNet34 & 85.70  & 0 \\
    ImageNet-ResNet50 & 75.84 & 0 \\
    GTSRB-VGG16 & 84.26 & -0.02 \\
    \bottomrule
    \end{tabular}%
  \label{tab:eval_model_util}%
\end{table}%

\noindent\textbf{Mitigating Untargeted Attacks.}  
We assess the effectiveness of \NAME in mitigating the untargeted BFA \cite{rakin2019bit}. 
The results in Table \ref{tab:eval_model_untarget} show a significant accuracy drop for models without \NAME's protection (the \textit{w/o. Def.} column). Notably, the ImageNet-ResNet50 model's accuracy is reduced to nearly zero. However, after applying \NAME (the \textit{w. Def.} column), all models demonstrate resilience, maintaining accuracy levels close to their original accuracy (\textit{Base} in the table). These results confirm that \NAME effectively mitigates untargeted attacks across a range of models.
\begin{table}[htbp]
  \centering
  \caption{Mitigating untargeted attacks.}
    \begin{tabular}{lrrrr}
    \toprule
    \multirow{2}[4]{*}{Model} & \multicolumn{1}{l}{\#Flipped} & \multicolumn{3}{c}{Acc (\%)} \\
\cmidrule{3-5}          & \multicolumn{1}{l}{Bits} & \multicolumn{1}{l}{Base} & \multicolumn{1}{l}{ w/o. Def.} & \multicolumn{1}{l}{w. Def.} \\
    \midrule
    ResNet20 & 30    & 85.36 & 11.46 & 82.53 \\
    ResNet32 & 14    & 84.38 & 11.81 & 82.00 \\
    ResNet34 & 30    & 85.70  & 19.84 & 83.63 \\
    ResNet50 & 10    & 75.84 & 0.10   & 75.85 \\
    VGG16 & 30    & 84.26 & 18.14 & 84.39 \\
    \bottomrule
    \end{tabular}%
  \label{tab:eval_model_untarget}%
\end{table}%

\noindent\textbf{Mitigating Targeted Attacks.}  
Targeted attacks differ from untargeted BFAs in that they seek to induce specific misclassifications without significantly impacting the overall model performance. 
To assess the effectiveness of \NAME in defending against targeted BFAs, we measure the attack success rate (ASR), which indicates the percentage of successful misclassifications made by the attacker.

We replicate the results using the open-sourced code and parameters recommended by the respective authors for each attack. Since TA-LBF functions on a per-image basis, we report the average number of flipped bits across images in the test dataset. Table~\ref{tab:eval_model_target} presents the ASR before (w/o. Def. column) and after applying \NAME (w. Def. column), along with the number of bits flipped during the attacks. \NAME significantly reduces the ASR across all three types of targeted attacks. Particularly, in high-complexity datasets like ImageNet, it reduces the ASR to near-zero for several attack types, underscoring its effectiveness in mitigating targeted misclassifications while preserving overall model accuracy.

\begin{table}[htbp]
  \centering
  \caption{Mitigating targeted attacks.}
  \resizebox{.85\linewidth}{!}{
    \begin{tabular}{clrrr}
    \toprule
    \multirow{2}[4]{*}{Attack} & \multirow{2}[4]{*}{Model} & \multicolumn{1}{l}{\#Flipped} & \multicolumn{2}{c}{ASR (\%)} \\
\cmidrule{4-5}          &       & \multicolumn{1}{l}{Bits} & \multicolumn{1}{l}{w/o. Def.} & \multicolumn{1}{l}{w. Def.} \\
    \midrule
    \multirow{5}[2]{*}{T-BFA} & ResNet20 & 6     & 99.84 & 8.16 \\
          & ResNet32 & 3     & 99.67 & 21.4 \\
          & ResNet34 & 20    & 99.83 & 10.43 \\
          & ResNet50 & 7     & 99.99 & 0.11 \\
          & VGG16 & 20    & 100   & 1.81 \\
    \midrule
    \multirow{5}[2]{*}{TBT} & ResNet20 & 97    & 93.99 & 2.22 \\
          & ResNet32 & 147   & 76.78 & 8.11 \\
          & ResNet34 & 130   & 77.49 & 6.04 \\
          & ResNet50 & 131   & 100   & 0 \\
          & VGG16 & 126   & 98.69 & 4.89 \\
    \midrule
    \multirow{5}[2]{*}{TA-LBF} & ResNet20 & 9.19  & 100   & 1.5 \\
          & ResNet32 & 10.92 & 100   & 1.9 \\
          & ResNet34 & 19.45 & 100   & 1.5 \\
          & ResNet50 & 12.76 & 100   & 0 \\
          & VGG16 & 18.42 & 100   & 2.7 \\
    \bottomrule
    \end{tabular}%
    }
  \label{tab:eval_model_target}%
\end{table}%

\noindent\textbf{Adaptive Attacks.}  
We consider a more sophisticated attacker who is aware of the obfuscation techniques used by \NAME, such as the insertion of dummy layers and neurons, but lacks knowledge of the specific quantities or exact locations of these modifications. He attempts to bypass \NAME's randomness in a brute-force manner, i.e., increasing the number of flipped bits. Rather than flipping only the identified vulnerable bits, the attacker targets a broader range of bits surrounding each vulnerable bit address \textit{addr} in an effort to counteract the effects of obfuscation. This range is defined as $[addr - x, addr + x]$, where $x$ determines the extent of the bit flips.

The results, as shown in Tables \ref{tab:eval_model_adaptive_1} and \ref{tab:eval_model_adaptive_2}, demonstrate that \NAME remains highly robust, against these adaptive attack strategies. In the case of untargeted adaptive attack (Table \ref{tab:eval_model_adaptive_1}), \NAME effectively maintains the model's accuracy, with no significant drop across various values of $x$. 
For targeted adaptive attack (Table \ref{tab:eval_model_adaptive_2}), \NAME also shows strong resistance. The attack success rate (ASR) remains remarkably low in most cases. 
These results confirm that \NAME's defense mechanisms, including randomizing bit locations and obfuscating vulnerable layers, are robust enough to withstand brute-force attacks, offering strong protection against both untargeted and targeted BFAs.

The robustness performance of \NAME can be attributed to the fact that most model parameters have a negligible impact on the overall output, even when flipped. Even if the attacker understands how \NAME operates, without precise knowledge of where the dummy operations are inserted, they can only rely on brute-force approaches to expand the flip range. Such brute-force methods barely succeed, as the increased range is less likely to target the obfuscated critical bits effectively.
Furthermore, targeted attacks generally require more precise bit selection and flipping compared to untargeted attacks. This makes that it is significantly harder for an attacker to execute a successful targeted attack without knowing the exact locations of the critical bits, explaining the minimal variation in the targeted attack results as the flip range expands. 
\begin{table}[htbp]
  \centering
  \caption{Evaluation for untargeted adaptive attack}
  \resizebox{.8\linewidth}{!}{
    \begin{tabular}{lrrrrr}
    \toprule
    \multirow{2}[4]{*}{Model} & \multicolumn{5}{c}{Acc (\%)} \\
\cmidrule{2-6}          & \multicolumn{1}{l}{$x$=3} & \multicolumn{1}{l}{$x$=5} & \multicolumn{1}{l}{$x$=7} & \multicolumn{1}{l}{$x$=9} & \multicolumn{1}{l}{$x$=11} \\
    \midrule
    ResNet20 & 76.11 & 80.20 & 82.14 & 82.14 & 82.14 \\
    ResNet32 & 75.05 & 76.62 & 76.24 & 75.97 & 75.64 \\
    ResNet34 & 85.29 & 82.41 & 81.73 & 83.32 & 83.13 \\
    ResNet50 & 74.36 & 75.84 & 75.83 & 75.83 & 75.87 \\
    VGG16 & 82.74 & 80.55  & 80.21 & 82.58 & 82.59 \\
    \bottomrule
    \end{tabular}%
    }
  \label{tab:eval_model_adaptive_1}%
\end{table}%

\begin{table}[htbp]
  \centering
  \caption{Evaluation for targeted adaptive attack}
  \resizebox{.85\linewidth}{!}{
    \begin{tabular}{clrrrrr}
    \toprule
    \multirow{2}[4]{*}{Attack} & \multirow{2}[4]{*}{Model} & \multicolumn{5}{c}{ASR (\%)} \\
\cmidrule{3-7}          &       & \multicolumn{1}{l}{$x$=3} & \multicolumn{1}{l}{$x$=5} & \multicolumn{1}{l}{$x$=7} & \multicolumn{1}{l}{$x$=9} & \multicolumn{1}{l}{$x$=11} \\
    \midrule
    \multirow{5}[2]{*}{T-BFA} & ResNet20 & 0     & 0     & 0     & 0     & 0 \\
          & ResNet32 & 1.83  & 1.83  & 1.83  & 1.83  & 1.83 \\
          & ResNet34 & 22.9 & 22.9 & 22.9 & 22.9 & 22.9 \\
          & ResNet50 & 0.11  & 0.11  & 0.11  & 0.11  & 0.11 \\
          & VGG16 & 1.55  & 1.55  & 1.55  & 1.55  & 1.55 \\
    \midrule
    \multirow{5}[2]{*}{TBT} & ResNet20 & 2.21  & 2.22  & 2.21  & 2.21  & 2.21 \\
          & ResNet32 & 8.14  & 8.13  & 8.15  & 8.12  & 8.18 \\
          & ResNet34 & 6.04  & 6.04  & 6.04  & 6.04  & 6.04 \\
          & ResNet50 & 0     & 0     & 0     & 0     & 0 \\
          & VGG16 & 4.89  & 4.89  & 4.89  & 4.89  & 4.89 \\
    \midrule
    \multirow{5}[2]{*}{TA-LBF} & ResNet20 & 1.5   & 0.99 & 1.98  & 1.98  & 1.98 \\
          & ResNet32 & 1.9   & 0.5   & 0.99  & 0.99  & 0.99 \\
          & ResNet34 & 1.5   & 1.49  & 2.97  & 1.98  & 1.98 \\
          & ResNet50 & 0     & 0     & 0     & 0     & 0 \\
          & VGG16 & 2.98  & 2.98  & 1.98  & 1.98  & 1.98 \\
    \bottomrule
    \end{tabular}%
    }
  \label{tab:eval_model_adaptive_2}%
\end{table}%

\begin{figure}[t]
    \centering
    \includegraphics[scale=0.55]{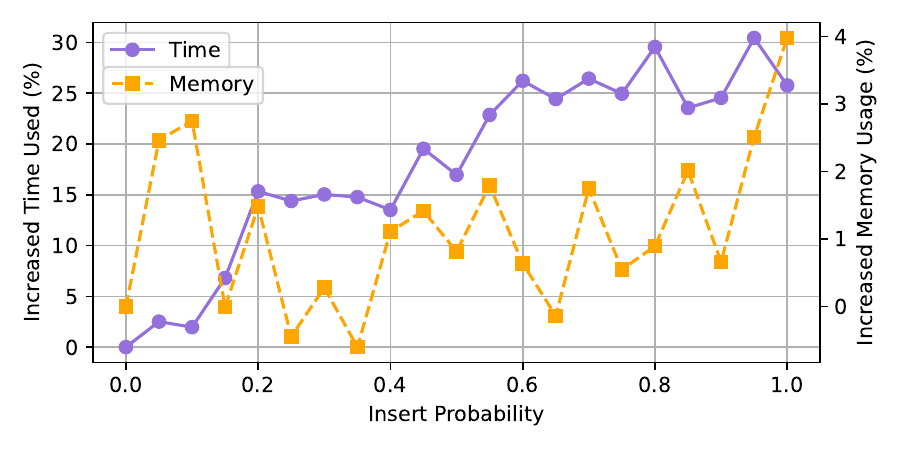}
    \caption{Time and memory Overhead.}
    \label{fig:eval_model_overhead}
\end{figure}

\noindent\textbf{Overhead.}  
The time and memory overhead incurred by \NAME is shown in Figure~\ref{fig:eval_model_overhead}, where "insert probability" denotes the likelihood of inserting dummy layers and neurons. We measure the percentage increase in both the time and memory required to run the CIFAR-10 test dataset using ResNet20, compared to an unobfuscated model. The reported values are averaged over 100 runs across all images in the test dataset. While \NAME incurs some time and memory overhead, it remains relatively low and acceptable. Notably, our observations suggest that an insert probability below 0.3 is sufficient in most cases, effectively minimizing the overhead without compromising the defense effectiveness.

\noindent\textbf{Comparison.} 
First, we evaluate \NAME against four untargeted defense schemes. The evaluation is performed using the ResNet20 model on CIFAR-10, and the results are presented in Table~\ref{tab:eval_model_untarget_compare}. 
We compare these defenses in terms of \textit{$\Delta$Acc (\%)} (change in model accuracy after applying the defense), \textit{$\Delta$Time (\%)} (execution time overhead introduced by the defense), \textit{$\Delta$Rounds (\%)} (increase in the number of attack rounds required for the attacker to achieve the same effect post-defense) and \textit{Mitigation Rate (\%)} (percentage of the attack successfully mitigated).
Since the source code for some of these schemes is unavailable, we rely on the data reported in their respective papers. The "-" symbol in the table indicates that a result was not reported in the corresponding paper.
We observe that \NAME introduces no accuracy loss (+0.01\%) and exhibits a moderate time overhead (+6.6\%). More notably, it significantly increases the number of attack rounds (160,831\%) required for successful bit-flip manipulation with 100\% mitigation rate, vastly outperforming all other methods in terms of resilience against untargeted BFAs. 
    
\begin{table}[htbp]
  \centering
  \caption{Comparison with untargeted defense schemes}
    \begin{tabular}{lllll}
    \toprule
    \multirow{2}[2]{*}{Defense} & \multicolumn{1}{l}{$\Delta$Acc} & \multicolumn{1}{l}{$\Delta$Time} & \multicolumn{1}{l}{$\Delta$Rounds} & \multicolumn{1}{l}{Mitigation} \\
    & \multicolumn{1}{l}{(\%)} & (\%)& (\%)  & Rate(\%)\\
    \midrule
    DeepShuffle & -     & 2.5  & 571 &-\\
    Neuropots & -1    & 9.7   & - & 93.3\\
    HammerDodger & -     & \textbf{0.8}   & - &-\\
    RREC  & \textbf{+0.03} & 5.8   & 1,452 &-\\
    \NAME & +0.01 &  6.6     &  \textbf{160,831}& \textbf{100}\\
    \bottomrule
    \end{tabular}%
  \label{tab:eval_model_untarget_compare}%
\end{table}%

Next, we compare \NAME with three targeted defense schemes. 
We adopt the ResNet32 model on CIFAR-10. The results are presented in Table~\ref{tab:eval_model_target_compare}, reporting the \textit{$\Delta$Acc (\%)} (the change in accuracy after applying the defense) and \textit{ASR (\%)} (the percentage of successful attacks in both TBT and TA-LBF attack scenarios). We observe that \NAME achieves the best performance, with a minimal accuracy drop (+0.01\%) and significantly lower ASR in both TBT (8.11\%) and TA-LBF (1.9\%) scenarios. This indicates that \NAME effectively reduces the success rate of targeted BFAs while better preserving accuracy than other defenses.

\begin{table}[htbp]
  \centering
  \caption{Comparison with targeted defense schemes}
    \begin{tabular}{lrrr}
    \toprule
    \multicolumn{1}{c}{\multirow{2}[4]{*}{Defense}} & \multicolumn{1}{c}{\multirow{2}[4]{*}{$\Delta$ Acc (\%)}} & \multicolumn{2}{c}{ASR (\%)} \\
    \cmidrule{3-4}          &       & \multicolumn{1}{l}{TBT} & \multicolumn{1}{l}{TA-LBF} \\
    \midrule
    BIN   & -2.26 & 94.8  & 100 \\
    RA-BNN & -1.71 & 74.5  & 100 \\
    Aegis & -1.26 & 19.9  & 6.3 \\
    \NAME & \textbf{+0.01} & \textbf{8.11} & \textbf{1.9} \\
    \bottomrule
    \end{tabular}%
  \label{tab:eval_model_target_compare}%
\end{table}%

\section{Conclusion}
\label{sec:Conclusion}
We introduce \NAME, an efficient and effective approach to mitigating the growing threat of BFAs targeting both model and code levels of DNN applications. By leveraging a novel obfuscation strategy that inserts randomized dummy operations, \NAME effectively complicates the attacker’s ability to locate and manipulate vulnerable bits, turning orchestrated attacks into random bit flips. Our evaluation shows \NAME's robustness against both untargeted and targeted BFAs across a range of datasets and architectures. It also brings minimal computational and storage overhead, making it a practical solution for real-world applications.

\bibliographystyle{plain}
\bibliography{mybib}{}
\end{document}